\documentclass[traditabstract]{aa} 
\usepackage{graphicx}
\usepackage{txfonts}
\usepackage{natbib} \bibpunct{(}{)}{;}{a}{}{,}
\usepackage{setspace}

\begin{document}
   \titlerunning{Orion KL: The hot core that is not a ``Hot Core''}
   \title{Orion KL: The hot core that is not a ``Hot Core''}

  \author{Luis A. Zapata\inst{1,2}, Johannes Schmid-Burgk\inst{1}
  \and Karl M. Menten\inst{1} }
  \authorrunning{Zapata et al.}    

  \institute{Max-Planck-Institut f\"{u}r Radioastronomie, Auf dem
  H\"ugel 69, 53121, Bonn, Germany \and Centro de Radioastronom\'\i a
  y Astrof\'\i sica, Universidad Nacional Aut\'onoma de M\'exico,
  Morelia 58090, M\'exico\\
  \email{lzapata@mpifr-bonn.mpg.de}}

  \date{Received ----; accepted  ----}

 \abstract{We present sensitive high angular resolution submillimeter
 and millimeter observations of torsionally/vibrationally highly
 excited lines of the CH$_3$OH, HC$_3$N, SO$_2$, and CH$_3$CN
 molecules and of the continuum emission at 870 and 1300 $\mu$m from the Orion
 KL region, made with the Submillimeter Array (SMA). These
 observations plus recent SMA CO J=3-2 and J=2-1 imaging of the explosive flow
 originating in this region, which is related to the non-hierarchical
 disintegration of a massive young stellar system, suggest that the
 molecular Orion ``Hot Core'' is a pre-existing density enhancement
 heated from the outside by the explosive event -- unlike in other
 hot cores we do not find any self-luminous submillimeter, radio or
 infrared source embedded in the hot molecular gas. 
 Indeed, we do not observe filamentary
 CO flow structures or ``fingers'' in the shadow of the hot core
 pointing away from the explosion center.  The
 low-excitation CH$_3$CN emission shows the typical molecular
 heart-shaped structure, traditionally named the Hot Core, and is
 centered close to the dynamical origin of the explosion. The highest
 excitation CH$_3$CN lines are all arising from the northeast lobe of
 the heart-shaped structure, {\it i. e.} from the densest and most highly
 obscured parts of the Extended Ridge.  The torsionally excited
 CH$_3$OH and vibrationally excited HC$_3$N lines appear to form a
 shell around the strongest submillimeter continuum source.
 Surprisingly the kinematics of the Hot Core and Compact Ridge
 regions as traced by CH$_3$CN and HC$_3$N also reveal filament-like
 structures that emerge from the dynamical origin. All of these
 observations suggest the southeast and southwest sectors of the
 explosive flow to have impinged on a pre-existing very dense part of
 the Extended Ridge, thus creating the bright Orion KL Hot Core. 
 However, additional theoretical and observational studies are required
 to test this new heating scenario. }

 \maketitle

\section{Introduction}

The famous eponymous hot core in the Orion Kleinmann-Low (KL)
star-forming region was given its name by \citet{Ho1979} who
identified it as a compact source of hot ammonia emission embedded in
a more extended ridge of dense material. A few years later,
\cite{Pauls1983}, using {\it the early Very Large Array (VLA)} at $2''$
resolution, showed the hot ammonia to arise from a heart-shaped region
of size $\approx 15''\times15''$ or $0.03\times0.03$~pc$^2$ at a
distance of 414 pc \citep{Mentenetal2007}.

Early millimeter interferometry with the Hat Creek Millimeter Interferometer (later
Berkeley-Illinois-Maryland Array, BIMA) and the Owens Valley Radio
Observatory (OVRO) Millimeter Array showed a chemically and
dynamically complex picture for the Orion hot core and (at least) one
other conspicuous region abutting it, the ``Compact Ridge''. Many
molecular lines arising from lower and moderately excited levels above
the ground state also showed the peculiar "heart" or "U" morphology,
for example CO, HCN, CH$_3$CN, H$^{13}$CN, HNCO, OCS, HDO, SO,
and SO$_2$ \citep{Gen1982, Mi1989, Wi1994,
Wri1996, bla1996, Che1996, Wil2000, Fri2008}. 
The LSR velocities had a centroid between 6 and 8 km
s$^{-1}$, close to the velocity of the ambient larger scale ridge
material ($\sim$ 9 km s$^{-1}$).

The heart-shaped structure occupies an area between two of the bona fide self-luminous 
sources in the region, source {\it I} and the Becklin-Neugebauer object ({\it BN}), 
which both showing weak radio emission.

Most of the molecules, including NH$_3$ but also C$_5$H$_3$CN,
CH$_2$H$_3$CN, (CH$_3$)$_2$CO, HCOOH, and HCOOCH$_3$, if imaged with
increasing resolution show clumpy structure with different morphology
but always well placed inside the area drawn by the heart-shaped structure
\citep{bla1996, Liu2002, Fri2008}. 
These molecular clumps show complex dynamics without a special trend
but with their emission's velocity centroid close to the systemic value
\citep{Maso1988, Wi1994, Liu2002, Fri2008}.

Soon after the first identification of the Orion hot core, hot ($\sim
$ 150 K), compact ($<0.05$ pc), dense ($10^6$~cm$^{-3}$) cores with
elevated abundances of many molecules were discovered in other
regions. The physical parameters that we give above for the hot cores
are not generalised, over the last decades has been found some other  
values. One of the first was found with the Hat Creek Interferometer
close (0.07 pc) to the archetypical ultracompact HII region W3OH
\citep{TurnerWelch1984}; see also
\citep{Mauersberger1986}. Subarcsecond resolution millimeter
interferometry with the IRAM Plateau de Bure Interferometer resolved
this ``Turner-Welch object'' into a protocluster consisting of three
sources \citep{Wyrowski1999}, one of which drives a powerful H$_2$O
maser outflow and two of which are associated with weak radio
continuum emission.  Many more hot cores with characteristics similar
to the above mentioned were found thereafter, all of which appear to
host one or more of the following: A hyper/ultra-compact HII region, a
strong millimeter source, a compact H$_2$O maser outflow and/or a
class II CH$_3$OH maser \citep[for reviews, see ][]{Kurtz2000,
  Cesaroni2005}. See for an example the compact hot cores associated
with intermediate mass protostars located just 0.15 pc south of KL in
Orion South \citep{zap2007}. Hot cores have even been found powered by
solar mass protostars: hot ''corinos'' \citep{Ceccarelli2000,
  Bottinelli2004}.

In contrast, what is heating the hot core in the Orion KL region and
whether, in particular, this core harbors (a) proto- or young stellar
object(s) has been the subject of debate for a long time.

\begin{table*}
\begin{minipage}[t]{\columnwidth}
\scriptsize
\renewcommand{\footnoterule}{}
\caption{Observational and physical parameters of submillimeter and millimeter lines}
\begin{center}
\begin{tabular}{lccccccc}
\hline \hline   
                & Rest frequency & E$_{lower}$/k & Range of Velocities
                & Linewidth\footnote{\scriptsize For the
                HC$_3$N(37-36) and CH$_3$OH(7$_{4,3}$-6$_{4,3}$)
                A$^-$ ($\nu_t$=1) torsionally/vibrationally excited lines the average
                linewidth given here is the sum of two velocity
                components, one centered at approximately $-$4 km
                s$^{-1}$ and the other one at $+$7 km s$^{-1}$.} &
                LSR Velocity & Peak Flux \\ Lines & [GHz] & [K] & [km
                s$^{-1}$] & [km s$^{-1}$] & [km s$^{-1}$] & [Jy
                Beam$^{-1}$] \\
\hline
\hline
CH$_3$CN(12$_3$-11$_3$) & 220.7090... & 122 & $-$10,$+$26 & 18 & 7 &
17 \\ CH$_3$CN(12$_6$-11$_6$) & 220.5944... & 315 & $-$5,$+$15 & 13 &
5 & 12 \\ CH$_3$CN(12$_9$-11$_9$) & 220.4039... & 636 & $+$2,$+$12 & 4
& 6 & 1 \\ CH$_3$OH(7$_{4,3}$-6$_{4,3}$) A$^-$ ($\nu_t$=1) &
337.9694... & 373 & $-$5,$+$15 & 5 & 8 & 13 \\
CH$_3$OH(7$_{4,3}$-6$_{4,3}$) A$^-$ ($\nu_t$=2) & 337.8775... & 705 &
$-$2,$+$11 & 4 & 8 & 3 \\ HC$_3$N(37-36)($\nu_7$=1) & 337.8240... &
613 & $-$12,$+$19 & 14 & 4 & 8 \\
SO$_2$(21$_{2,20}$-21$_{1,21}$)($\nu_2$=1) & 337.8925... & 948 &
$-$1,$+$13 & 5 & 6 & 2 \\
\hline \hline
\end{tabular}
\end{center}
\end{minipage}
\end{table*}

\citet{bla1996}, using line and continuum observations with OVRO,
found no evidence for a luminous internal heating source within the
Orion KL Hot Core. BIMA observations of formic acid (HCOOH) and the
positions of H$_2$O masers \citep[see {\it e.g.}][]{Gen1981,Gau1998} in 
the Compact Ridge region suggested that these molecules delineate the
interaction region between the Orion KL outflow and the ambient
quiescent gas \citep{Liu2002}.
\citet{Che1996} found that their CO data were consistent with a 
northwest-southeast biconical outflow centered close to the positions
of {\it BN} and source {\it I} that is partly truncated by the Hot Core.  This is
quite suggestive of the outflow's energy being partly dissipated in 
heating the Orion KL Hot Core.

On the other hand, \citet{Kau1998} proposed the Orion KL Hot Core to
more likely be heated by stars embedded within the core rather than
powered from outside because of the core's large column densities
(N(H$_2$) $\geq$ 10$^{23}$ cm$^2$) and warm temperatures (T $\geq$ 100 K).
Likewise to them the distribution of vibrationally 
excited HC$_3$N emission suggested that the Orion hot core was heated 
from inside by a group of stars \citet{de2002}.

A group of strong near- and mid-infrared sources have been proposed to
be responsible for heating the Orion KL Hot Core and the Compact Ridge
\citep{Gen89, bla1996}. However, sensitive VLA radio observations made
by \citet{men1995} revealed that most of the IR sources in the Orion
KL nebula are not self-luminous but rather show reprocessed emission escaping
through inhomogeneities in the dense material.
In particular, these data showed that no part of the  ''IRc2'' group of IR sources 
was coincident with the conspicuous radio continuum source {\it I} 
which lies within the boundaries but not at the center
of the heart-shaped molecular structure.  Furthermore, OVRO and
CARMA millimeter continuum maps by \citet{bla1996,Fri2008}
demonstrated that the millimeter and IR sources do not show good
correspondence.  \citet{bla1996, Wri1996, Cha1997} in addition found
that the dust and the peak molecular emission coincide neither with
Source {\it I} nor with IRc2, the radio and infrared sources closest to
the Hot Core.

In this paper, we present sensitive high angular resolution
submillimeter and millimeter observations of the Orion KL region that
were made in an attempt to understand the nature of the hot molecular
material it harbors. We used the Submillimeter Array to image the
continuum at 870 and 1300 $\mu$m and a series of molecular lines emitted from
energy levels with moderate, high, and very high energies (122--948 K)
above the ground state, namely CH$_3$OH in the first and the
second torsionally excited states, HC$_3$N and SO$_2$ in
vibrationally excited states and high excitation lines from the
vibrational ground state of CH$_3$CN.  

In Section 2 we discuss the observations, in
Section 3 we present and discuss our SMA millimeter and submillimeter
data. Finally, in Section 4 we give the main conclusions drawn 
from the observations.

\section{Observations}

\subsection{Millimeter}

Observations were made with the Submillimeter Array\footnote{The
Submillimeter Array is a joint project between the Smithsonian
Astrophysical Observatory and the Academia Sinica Institute of
Astronomy and Astrophysics, and is funded by the Smithsonian
Institution and the Academia Sinica.} (SMA) during 2007 January and
2009 February. The SMA was in its compact and sub-compact
configurations with baselines ranging in projected length from 6 to 58
k$\lambda$. We used the mosaicking mode with half-power point spacing 
between field centers and covered the entire Orion Hot Core.
The primary beam of each pointing at 230 GHz has a FWHM
diameter of about 50$''$.

The receivers were tuned to a frequency of 230.5387970 GHz in the
upper sideband (USB), while the lower sideband (LSB) was centered on
220.5387970 GHz. The CH$_3$CN(12$_k$-11$_k$) $k$-ladder with $k = 10,
\ldots,1, 0$ was detected in the LSB at frequencies around of 220.3 --
220.7 GHz. See Table 1 for their rest frequencies.  
A full astrochemical analysis of the data on CH$_3$CN (and
other species) is beyond the scope of the present paper in which we
concentrate on the spacial distributions of three lines of high and very
high excitation, namely $k =3/E_{\rm l} = 122 {\rm K}$, $k =6/E_{\rm
l} = 315 {\rm K}$, $k =9/E_{\rm l} = 636 {\rm K}$.

The full bandwidth of the SMA correlator is 4 GHz (2 GHz in each
band).  The SMA digital correlator was configured in 24 spectral
windows (``chunks'') of 104 MHz each, with 256 channels distributed
over each spectral window, thus providing a spectral resolution of
0.40 MHz (0.54 km s$^{-1}$) per channel. However, in this study we
smoothed the spectral resolution to about 1 km s$^{-1}$.

The zenith opacity ($\tau_{230 GHz}$) was $\sim$ 0.1 -- 0.3,
indicating reasonable weather conditions.  Observations of Uranus and
Titan provided the absolute scale for the flux density calibration.
Phase and amplitude calibrators were the quasars 0530+135, 0541-056,
and 0607-085. The uncertainty in the flux scale is estimated to be
15-20$\%$, based on the SMA monitoring of quasars.  Further technical
descriptions of the SMA and its calibration schemes can be found in
\citet{Hoetal2004}.

The data were calibrated using the IDL superset MIR, originally
developed for the Owens Valley Radio Observatory
\citep{Scovilleetal1993} and adapted for the SMA.\footnote{The MIR
cookbook by C.  Qi can be found at
http://cfa-www.harvard.edu/$\sim$cqi/mircook.html} The calibrated data
were imaged and analyzed in standard manner using the MIRIAD, GILDAS
and AIPS packages.  We used the ROBUST parameter set to 0 to obtain an
optimal compromise between sensitivity and angular resolution.  The
line image rms noise was around 200 mJy beam$^{-1}$ for each channel
at an angular resolution of $3\rlap.{''}28$ $\times$ $3\rlap.{''}12$
with a P.A. = -14.0$^\circ$.

\begin{figure*} 
\begin{center}
\includegraphics[scale=0.79,angle=-90]{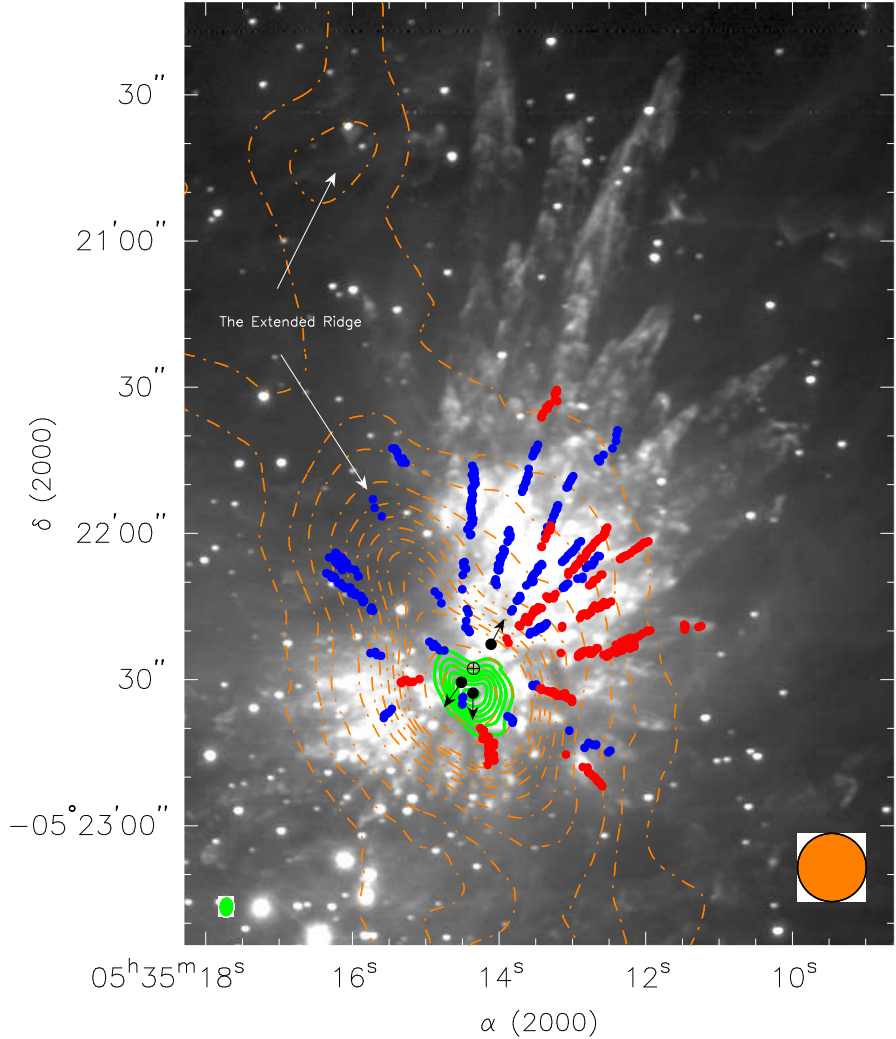} 
\caption{H$_2$ image of the Orion KL region (Bally et al. 2010, in preparation)
overlaid with an integrated intensity SMA map of 
CH$_3$CN(12$_3$-11$_3$) (green contours), the positions of the
blueshifted and redshifted CO(2-1) fingers from \citet{Zapataetal2009}
(blue and red dots, respectively) and a SCUBA 850 $\mu$m continuum map
from \citet{John1999} (brown dashed contours). The green contours are -2,
2, 4, 6, 8, 10, 12, 14, 16, 18, and 20 times 30 Jy beam$^{-1}$ km
s$^{-1}$, the rms noise of the image. The brown dashed contours are -2,
2, 4, 6, 8, 10, 12, 14, 16, 18, and 20 times 3.5 Jy beam$^{-1}$. 
The black circles with vectors mark the
positions and orientations of the proper motions of the radio and
infrared sources {\it BN}, {\it I} and n
\citep{Rodriguezetal2005,Gomezetal2005}.
The black open circle with a cross represents the zone from where the three sources
were ejected some 500 years ago, and the origin of the Orion KL
molecular outflow as suggested by SMA CO(2-1) observations
\citep{Gomezetal2005, Zapataetal2009}. The beam size of the SMA
CH$_3$CN(12$_3$-11$_3$) and the SCUBA maps are shown at the left- and
right-hand bottom, respectively. The
synthesized beam size of the SMA is $3\rlap.{''}28$ $\times$
$3\rlap.{''}12$ with a P.A. = -14.0$^\circ$. The beam size of the
SCUBA submillimeter observations is about 14$''$.}
\label{Fig1} 
\end{center} 
\end{figure*}

\subsection{Submillimeter}

Data collected on 2003 December 12 were retrieved from the SMA archive. 
At the time of these observations the SMA had seven antennas
in its compact configuration with baselines
ranging in projected length from 30 to 255 k$\lambda$.  The primary
beam of the SMA at 345 GHz has a FWHM of about 30$''$. The molecular
from the hot core was found well inside of the primary beam.

The receivers were tuned to a frequency of 347.33056 GHz in the upper
sideband (USB), while the lower sideband (LSB) was centered on
337.33056 GHz.  The LSB contained high excitation rotational lines from the 
first and second torsionally states of CH$_3$OH $\nu_t$=1 and 2, 
respectively and from vibrationally excited states of
HC$_3$N ($\nu_7$=1) and SO$_2$ ($\nu_2$=1); 
see Table 1 for their rest frequencies and lower 
level energies above the ground state, $E_{\rm l}$.  
The SMA digital correlator was
configured in 24 spectral windows (``chunks'') of 104 MHz each, with
128 channels distributed over each spectral window, thus providing a
spectral resolution of 0.81 MHz (0.72 km s$^{-1}$) per channel.
However, we smoothed the spectral resolution to 1.0 km s$^{-1}$
per channel.

\begin{table*}[ht]
\begin{minipage}[t]{\columnwidth}
\renewcommand{\footnoterule}{}
\scriptsize
\caption{Observational parameters of the submillimeter 
        and millimeter compact sources}
\begin{center}
\begin{tabular}{lccccccc}
\hline \hline   
                & & & \multicolumn{2}{c}{870 $\mu$m} &
                \multicolumn{2}{c}{1300 $\mu$m} \\ \hline & RA & DEC
                &  Flux Density & Peak Flux &  Flux Density & Peak Flux
                & Spectral \\ Sources & J[2000] & J[2000] & [Jy] &
                [Jy Beam$^{-1}$] & [Jy] & [Jy Beam$^{-1}$] &
                Index\footnote{\scriptsize The beam size of the two
                measurements is quite similar, thus {\it warranting that we} 
                are sensitive to the same spatial scales. }  \\
\hline
\hline
SMM1 & 5 35 14.015 & -5 22 36.88 & 4.0 & 1.5 & 1.3 & 0.7 & 2.9 \\
SMM2 & 5 35 14.087 & -5 22 27.55 & 5.6 & 2.6 & 1.5 & 0.7 & 3.3 \\ 
SMM3 & 5 35 14.560 & -5 22 31.38 & 11.1 & 4.2 & 4.8 & 2.7 & 2.1 \\
\hline \hline
\end{tabular}
\end{center}
\end{minipage}
\end{table*}

The zenith opacity ($\tau_{230 GHz}$) was $\sim$ 0.035 -- 0.04,
indicating excellent weather conditions.  Observations of Uranus and
Callisto provided the absolute scale for the flux density calibration.
Phase and amplitude calibrators were the quasars 0420$-$014, and 3C120.
The uncertainty in the flux scale is also estimated to be 15--20$\%$,
based on the SMA monitoring of quasars.

The calibrated data were imaged and analyzed in standard manner using
the MIRIAD, and AIPS packages. We used the ROBUST parameter set to 2
to obtain a slightly better sensitivity while losing some angular
resolution.  The line image rms noise was around 170 mJy beam$^{-1}$
for each channel at an angular resolution of $4\rlap.{''}2$ $\times$
$1\rlap.{''}30$ with a P.A. = $-42.0^\circ$.

These archive data had already been presented in \citet{buu2005}. However, we found
important differences between our images of the vibrationally excited 
lines and the higher resolution images of
that paper. This is discussed in more detail in the
next section.

\section{Results and Discussion}

\subsection{Evidence of a close relation between the Orion KL Hot Core 
and the explosive disintegration event}

In the following we shall present evidence that the
Orion KL Hot Core is not at all a typical hot core, with internal heating by
a nascent star, but rather a condensation energized by the impact of 
high velocity material that originates some 0.01 pc away in an external 
star forming concentration.

Figure 1 shows an overlay of the H$_2$ ``fingers'' (Bally et al. 2010,
in prep.) the blue- and red-shifted CO(2-1) filaments
\citep{Zapataetal2009}, the integrated intensity distribution of the lower excitation 
CH$_3$CN (12$_3$-11$_3$) line, the positions of the three runaway
stars ({\it BN}, {\it I} and n) and their dynamical center
\citep{Rodriguezetal2005,Gomezetal2005}, and the submillimeter dust
emission from the Orion KL region as mapped by the Submillimeter
Common Bolometer Array (SCUBA) with the James-Clerk-Maxwell Telescope
\citep{John1999}.  The intensity of the CH$_3$CN (12$_3$-11$_3$) line
is integrated over the whole broad velocity range it covers, i.e. from
$-$10 to $+$26 km s$^{-1}$. This thermal emission shows the classical
heart-shaped morphology first revealed in $2''$ resolution NH$_3$ images
by \citet{Pauls1983} and later observed in many millimeter lines
\citep[see, e.g. ][]{Wi1994} . The heart is centered to the south of
the dynamical origin of the three runaway stars and of the CO
filaments, approximately in the middle between the positions
of source {\it I} and {\it BN} \citep{Zapataetal2009}. The eastern lobe of the
heart-shaped molecular structure shows a slight extension 
towards NNE with the same orientation as the highly obscured
Extended Ridge as traced by the SCUBA observations. One clearly sees a
correlation between the dark areas of the H$_2$ image and the
submillimeter emission from the Extended Ridge, revealing the high
extinction of this region.

Figure 1 shows that the northwest CO filaments are much better defined
and longer than those located toward the southeast and southwest side
of Orion KL where the Hot Core is located. Also the H$_2$ fingers
located in the northwest quadrant seem to be brighter
and stronger than those found to the South. This suggests a
possible relationship between the position of the Orion KL Hot Core
and the absence of strong filaments in the described directions.

\begin{figure*} 
\begin{center}
\includegraphics[scale=0.55,angle=-90]{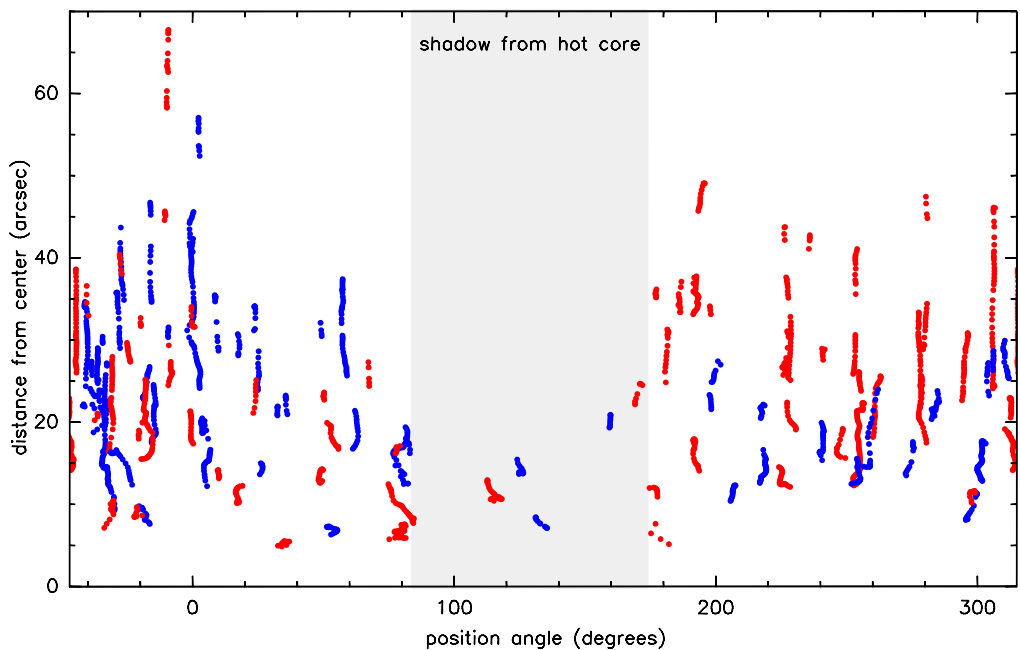} 
\caption{Distance from the explosion origin or center vs. position 
angle on the sky of the CO(3-2) molecular filaments obtained by Zapata et
al. (2010, in prep.). The explosive origin is located in $\alpha[2000]$=05$^h$ 35$^m$ 14.37$^s$ and $\delta[2000]$=-05$^\circ$ 22$'$ 27.9$''$.
The light grey area marks the ``shadow''
caused by the Orion KL Hot Core; filaments are prevented by the Hot Core from expanding into
this sector. The blue filaments indicate the CO(3-2) molecular gas approaching 
towards us, while red ones receding.}
\label{Fig2} 
\end{center} 
\end{figure*}

In order to gain better insight into this relationship we show in Figure 2
a diagram of the distance from the explosive origin vs. position angle on the
sky of the CO$(3-2)$ filaments discovered by Zapata et
al. (2010, in prep.). The CO(2-1) filaments found by
\citet{Zapataetal2009} and shown in Figure 1 are very similar to these
but somewhat less well-defined.  The
diagram demonstrates how closely the absence of CO
filamentary structures corresponds to the shadow area behind
the Hot Core as observed in CH$_3$CN $(12_3-11_3)$ emission.  This
suggests that the bulk of the high velocity CO gas
accelerated by the explosion that travels in these directions was stopped
by a high density cloud ({\it i.e.} the Extended Ridge).
The cloud was hereby heated, creating in the process
the Orion KL Hot Core.

As mentioned above, the strong water maser emission observed toward
the Hot Core region \citep{Liu2002} as displayed in Figure 3 supports
this picture, as probably does the OH maser emission. Interstellar OH
masers exist mostly in the slowly (few km~s$^{-1}$) expanding, warm
(150 K), dense ($10^{6-7}$~cm$^{-3}$) envelopes of ultracompact HII
regions \citep{Bloemhof1996,Fish2005}, but weak ones have recently
been found as well, associated with the H$_2$O outflow from the W3OH TW
object \citep{Argon2003}. We note that the total velocity spread of
the Orion KL OH masers ($\approx 55$~km~s$^{-1}$) is about 10 times
bllarger than that found for a typical interstellar OH maser source
\citep{Co2006}. The latter reference describes the idiosyncratic
behavior of the emission in the different OH hyperfine
transitions.

\begin{figure*} 
\centering 
\includegraphics[scale=0.32]{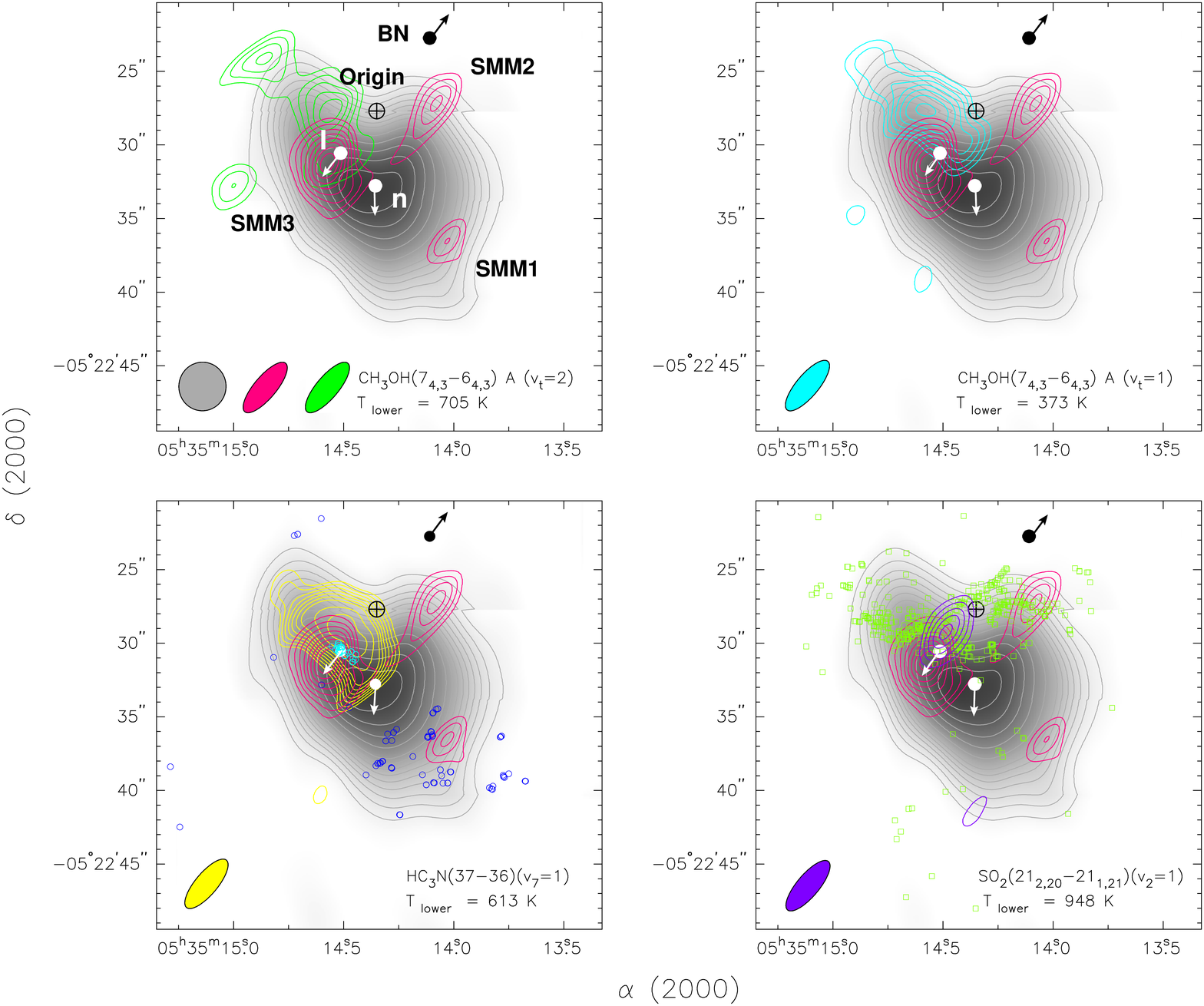}
\caption{Overlays of the SMA 870 $\mu$m continuum map  (pink contours)
and the integrated line emission maps of four submillimeter
tosionally/vibrationally excited lines towards the Orion {\it BN}-KL region
onto the CH$_3$CN(12$_3$-11$_3$) integrated line emission map (grey).
The pink contours are from 40\% to 90\% with steps of 5\% of
the peak of the continuum emission, 4.5 Jy
beam$^{-1}$. The grey contours are -2, 2, 3, 4, 5, 6, 7, 8, 9, 10, 12, 14, 16, 18, and 20
times 30 Jy beam$^{-1}$ km s$^{-1}$, the rms noise of the image. 
Upper left panel:  CH$_3$OH(7$_{4,3}$-6$_{4,3}$) A$^-$
($\nu_t$=2): green contours. The contours
are from 35\% to 90\% with steps of 10\% of the peak of the line
emission, 37 Jy beam$^{-1}$ km s$^{-1}$. Upper right
panel:  CH$_3$OH(7$_{4,3}$-6$_{4,3}$) A$^-$ ($\nu_t$=1)
(blue contours).  The contours are from 35\%
to 90\% with steps of 10\% of the peak of the line emission,
50 Jy beam$^{-1}$ km s$^{-1}$.  Lower left panel:  
HC$_3$N(37-36)($\nu_7$=1) (yellow contours).
The contours are from 20\% to 90\% with steps of 10\% of the peak of
the line emission, 170 Jy beam$^{-1}$ km s$^{-1}$. 
Lower right panel: SO$_2$(21$_{2,20}$-21$_{1,21}$)($\nu_2$=1)
(purple contours).  The contours are from
40\% to 90\% with steps of 5\% of the peak of the integrated emission,
35.1 Jy beam$^{-1}$ km s$^{-1}$. In every panel the black
and white circles with vectors mark the position and the orientation
of the proper motions of the radio and infrared sources {\it BN}, {\it I}, and {\it n}.
The black open circle with a cross represents the zone from where the three sources
were ejected some 500 years ago and which is the origin of the
Orion-KL molecular outflow as suggested by the SMA CO(2-1)
observations. The synthesized beams of the SMA millimeter and
submillimeter observations are shown in the bottom left hand part of
each panel. The blue open circles and green squares mark the positions
of the H$_2$O and OH maser spots, respectively \citep{Gau1998,Co2006}.
The light blue open circles show the water masers close to Source {\it I}.}
\label{Fig3} 
\end{figure*}

\subsection{Distribution and kinematics 
of the molecular gas within the Orion KL Hot Core}

\subsubsection{Torsionally/vibrationally highly excited lines}

In Figure 3, we show the images of the distributions of four very
torsionally/vibrationally highly excited lines,
CH$_3$OH(7$_{4,3}$-6$_{4,3}$) A$^-$ ($\nu_t$=2),
CH$_3$OH(7$_{4,3}$-6$_{4,3}$) A$^-$ ($\nu_t$=1),
HC$_3$N(37-36)($\nu_7$=1), and
SO$_2$(21$_{2,20}$-21$_{1,21}$)($\nu_2$=1), and the
870 $\mu$m continuum, overlaid on the moderate excitation
CH$_3$CN ($12_3-11_3$) line toward the core of the Orion KL region. These
four lines have energies up to 943 K and thus trace the
hottest molecular material. The continuum emission arises
from only three compact sources that were already reported at
millimeter wavelengths by \citet{bla1996, Wri1996, pla1995}; namely
the Hot Core -$>$ SMM3, the Northwest Clump -$>$ SMM2, and the Compact
Ridge -$>$ SMM1. In Table 2, we list their observational
parameters. The source SMM1 is further resolved into three
compact sources: SMA 1, Hot Core, and Source {\it I} \citep{idio2004}.  The
1300 $\mu$m continuum emission is quite similar to that at 870
$\mu$m so that we do not show a map of this wavelength.

The torsionally/vibrationally highly excited emission is seen to be
compact and located exclusively on the northeast side of the heart-shaped
structure as traced by the emission of CH$_3$CN(12$_3$-11$_3$). We list
the observational and physical parameters of these
lines in Table 1. Clearly this CH$_3$OH and
HC$_3$N emission shows pronounced
extensions or tails toward the northeast part of the ,
oriented in the direction of the Extended Ridge.  Such a tail or extension is
also observed in CH$_3$CN(12$_3$-11$_3$) as already mentioned in the
previous section.  The HC$_3$N(37-36)($\nu_7$=1) and
CH$_3$OH(7$_{4,3}$-6$_{4,3}$) A$^-$ ($\nu_t$=2) line emissions
seem to surround the submillimeter source SMM3 like
part of a shell, and both lines peak between that submillimeter source and the
origin of the explosion. The morphology of the HC$_3$N(37-36)($\nu_7$=1)
emission even appears to point toward this origin.  The
vibrationally excited SO$_2$ emission, with an energy of the lower state of
almost 1000 K (see Table 1), is very compact and peaks at the same
position as does HC$_3$N(37-36)($\nu_7$=1) emission.

\begin{figure*} 
\centering 
\includegraphics[scale=0.35]{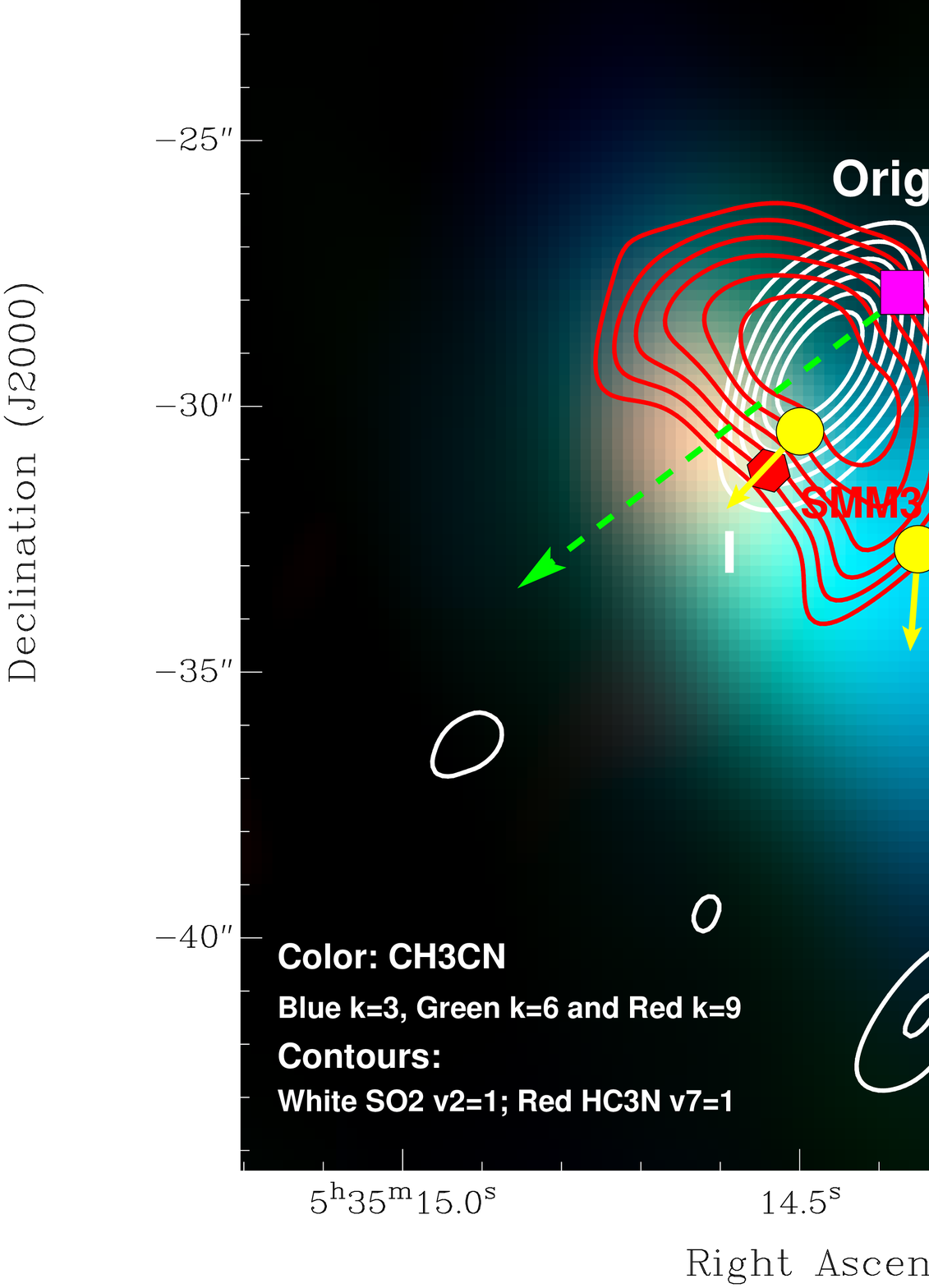}
\caption{  SMA CH$_3$CN(12$_k$-11$_k$) composite image with red 
          representing the rotational transition $k$=$9$, green
          $k$=$6$, and blue $k$=$3$. Overlaid on this image are
          HC$_3$N(37-36)($\nu_7$=1) and
          SO$_2$(21$_{2,20}$-21$_{1,21}$)($\nu_2$=1) integrated
          emissions (red and white contours, respectively).  The red
          contours are from 30\% to 90\% with steps of 12\% of the
          peak of the line emission (170 Jy beam$^{-1}$ km s$^{-1}$).
          The integrated velocity range of the
          HC$_3$N(37-36)($\nu_7$=1) line emission is from $-$12 to
          $+$19 km s$^{-1}$.  The white contours are from 40\% to 90\%
          with steps of 5\% of the peak of the integrated emission
          (35.1 Jy beam$^{-1}$ km s$^{-1}$). The integrated velocity
          range of the SO$_2$(21$_{2,20}$-21$_{1,21}$)($\nu_2$=1) line
          emission is from $-$1 to $+$13 km s$^{-1}$. The yellow
          circles with vectors mark the same positions as in Figures 3 and 4,
          the pink square represents the origin of
          the Orion KL molecular outflow as suggested by our SMA
          CO(2-1) observations.  The red hexagons mark the positions
          of the three submillimeter sources SMM1, SMM2, and SMM3,
          the green dashed arrow the position and orientation of
          the filament structure revealed by the moment two map of
          HC$_3$N(37-36)($\nu_7$=1), see Figure 5.  }
\label{Fig4} 
\end{figure*}

It is important to note that the vibrationally/torsionally
excited lines are not associated with any radio or submillimeter
continuum sources and always fall well outside or on the edge of the
heart-shaped structure as traced by CH$_3$CN(12$_3$-11$_3$); see Figure 3.
This suggests that the heating source of the Hot Core
is external as proposed by \citet{bla1996} and \citet{Liu2002}.
However, here we refer ``external heating'' to the input 
of mechanical energy
from the explosive disintegration to the dusty core rather than the input 
of radiative energy from an external stellar source located at a 
certain distance.
 
If the heating of the Orion KL Hot Core were internal one might
expect the hot molecular gas to be centered on a strong compact
continuum source located within the core, but this is not the case.
The fact that both HC$_3$N(37-36)($\nu_7$=1) and
CH$_3$OH(7$_{4,3}$-6$_{4,3}$) A$^-$ ($\nu_t$=2) seem to form a shell
around the continuum source Hot Core speaks against any internal
heating scenario, rather suggesting that the heating to come from outside.

The torsionally/vibrationally highly excited lines are exclusively placed
along the densest and most highly obscured parts of Orion KL, i.e. on the
Extended Ridge. We do not find any such emission associated with the
Compact Ridge or the Northwest Clump. This is in very good agreement
with the observations of vibrationally excited HC$_3$N
\citep{bla1996, Wri1996,de2002}. However, comparison of our 
submillimeter line maps with those of millimeter wavelengths further reveals
that the hottest molecular gas is only located in the northwest part of the
millimeter/submillimeter source SMM3.

The high angular resolution maps ($\leq$ 1$''$) presented in
\citet{buu2005} differ considerably from ours as well as from the
vibrationally excited line emission presented by \citet{bla1996,
Wri1996,de2002} at millimeter wavelengths.  In order to understand these
differences between the SMA maps we have recalibrated the high
angular resolution data (a factor of more than four
better than our actual resolution) of Beuther et al. and made some images of their
vibrationally/torsionally excited lines.  In all these maps we found
pronounced negative and positive sidelobe structures going from west
to east, with most of the compact positive structures showing nearly
random behavior that prohibits us from determining which of these
compact sources are real. This suggests that part of the emission
here must have been resolved out.

\subsubsection{Emission of CH$_3$CN(12-11) in the vibrational ground state}

In Figure 4, we show an SMA CH$_3$CN(12$_k$-11$_k$) color composite of
the rotational transitions $k$=$9$, $k$=$6$ and $k$=$3$
overlaid with the HC$_3$N(37-36)($\nu_7$=1)
and SO$_2$(21$_{2,20}$-21$_{1,21}$)($\nu_2$=1) integrated line
emissions.  The image reveals that CH$_3$CN(12$_9$-11$_9$), which is
supposed to trace positions of very hot molecular gas (see
Table 1), is also located in the northeast part of the heart-shaped 
structure where all
the torsionally/vibrationally excited lines are found. This confirms
that the hottest molecular gas is found in this part of
the Orion KL Hot Core. However, there is a clear offset (toward the
southeast) between the vibrationally/torsionally excited lines and this
ground state line. The CH$_3$CN(12$_3$-11$_3$) emission, on the other hand, reveals
that the coldest part of the molecular heart-shaped structure is located
toward the Compact Ridge.

\begin{figure*} 
\centering 
\includegraphics[scale=0.37]{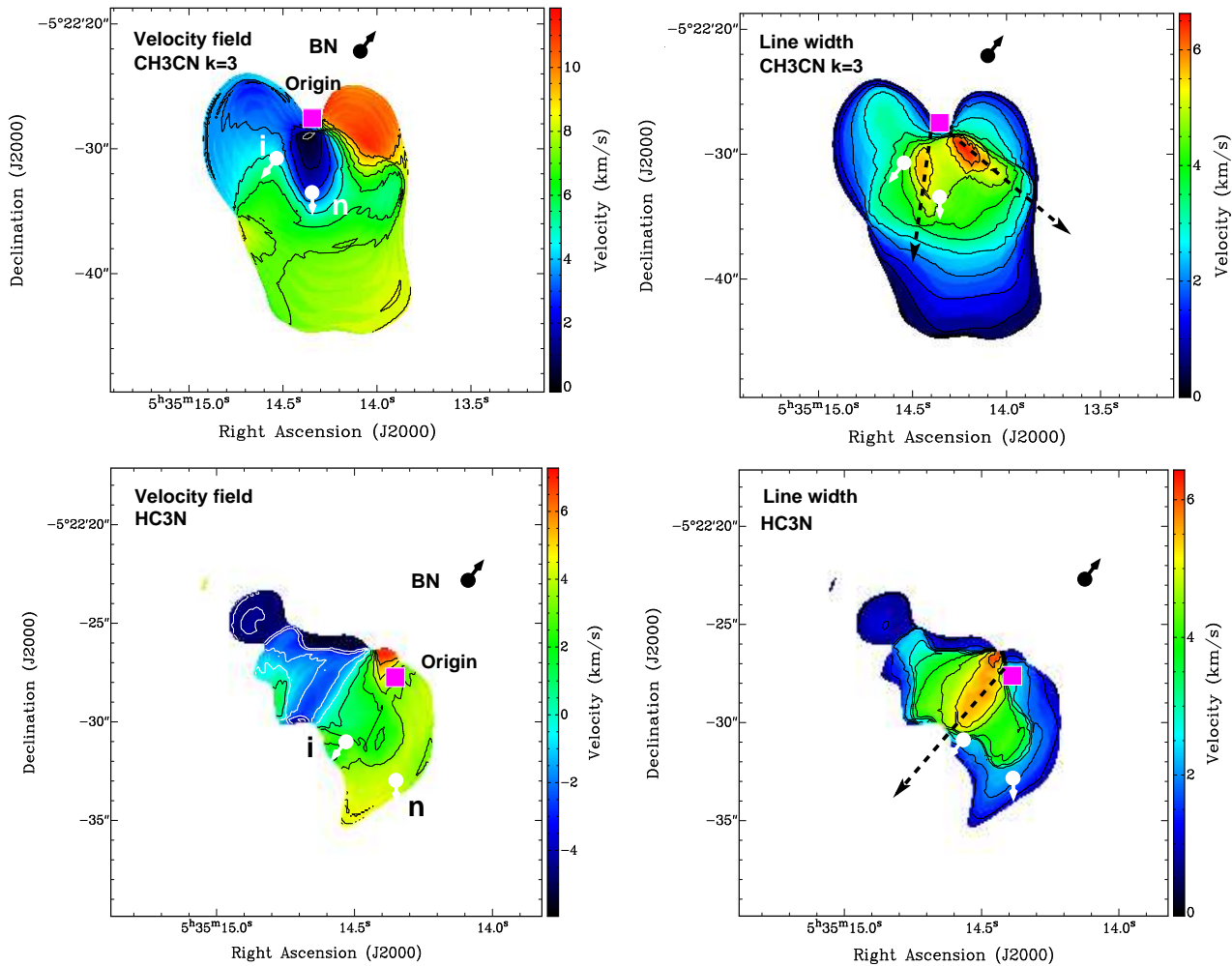}
\caption{SMA moment one and two color scale maps of the emission 
of CH$_3$CN(12$_3$-11$_3$) (upper images) and
HC$_3$N(37-36)($\nu_7$=1) (lower images), respectively.  The black and
white circles are as in Figure 3. The violet square represents the
origin of the CO(2-1) filaments.
\citep{Zapataetal2009}.}
\label{Fig5} 
\end{figure*}

All the emission peaks of these lines,
(CH$_3$CN(12$_9$-11$_9$), SO$_2$(21$_{2,20}$-21$_{1,21}$)($\nu_2$=1),
and HC$_3$N(37-36)($\nu_7$=1)), are well concentrated in the southeast
sector of the filamentary CO structure 
and are aligned with the band of highest HC$_3$N(37-36)($\nu_7$=1) 
line width, see Figure 5.

Note that the ground state line CH$_3$CN(12$_9$-11$_9$) is not
coincident with any of the submillimeter or radio
continuum sources located in this region.

\subsubsection{Kinematics}

We show the kinematics of the molecular gas within the Hot Core and
the Compact Ridge regions in Figure 5 where
moment one (integrated weighted velocity) and two (integrated velocity
dispersion squared) maps of the vibrationally excited resp. ground state
emission from the CH$_3$CN(12$_3$-11$_3$) and
HC$_3$N(37-36)($\nu_7$=1) lines are displayed.  The moment one maps show
that the blueshifted molecular gas is concentrated in the
northeast part of the heart-shaped structure, while the redshifted
gas is located on the northwest part. The origin of the runaway stars
and of the filamentary structure is well placed in between, in the middle of the
northern lobes of the heart-shaped structure as defined by
CH$_3$CN(12$_3$-11$_3$). The emission of HC$_3$N(37-36)($\nu_7$=1) is
totally blueshifted with respect to ambient and shows a velocity
gradient with northeast-southwest orientation. 

The moment two maps surprisingly reveal filamentary structures that clearly point
towards the dynamical origin.  These filaments are reminiscent of the 
filamentary CO(2-1) emission found by \citet{Zapataetal2009}
that might have been caused by the ejection of  material
upon the dynamical disintegration of the young stellar system {\it BN}, {\it I},
and {\it n}.  Note that the largest linewidths are located closest
to this origin.

We think that maybe the $-$4 km s$^{-1}$ blushifted component 
found in HC$_3$N(37-36)($\nu_7$=1) and CH$_3$OH(7$_{4,3}$-6$_{4,3}$) 
is excited for one filament that collided with a slightly different velocity 
toward this position.

\begin{figure*} \centering \includegraphics[scale=0.3]{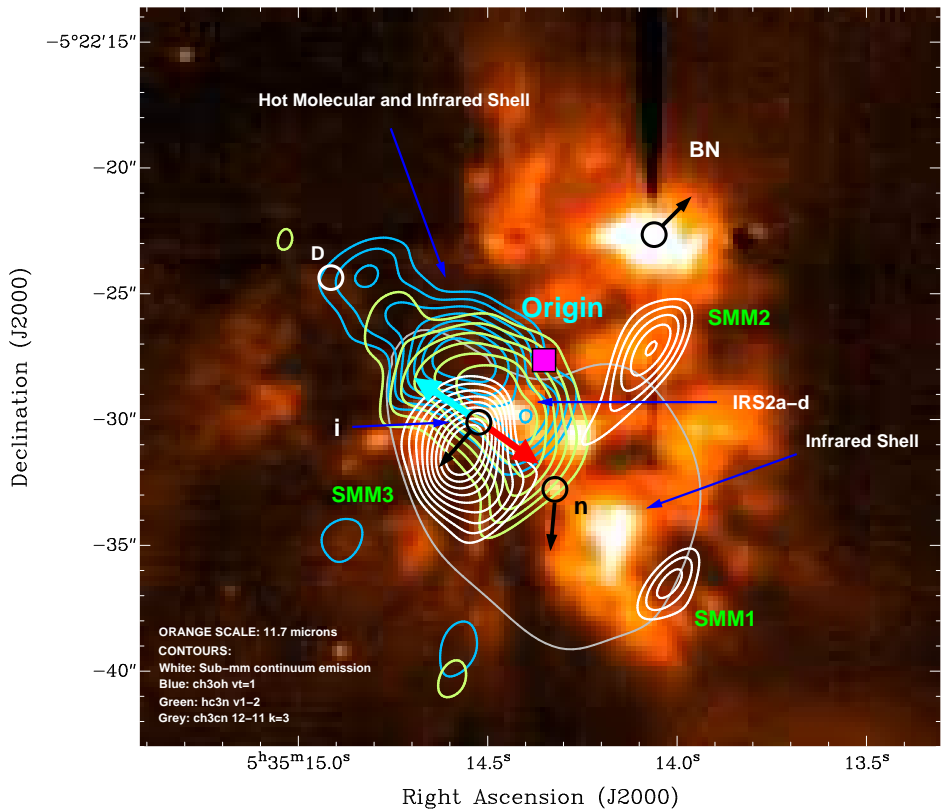}
\caption{Gemini  11.7 $\mu$m continuum image (orange) from
\citet{smith2005} overlaid with SMA maps of the submillimeter
continuum emission (faint white contours), the
CH$_3$OH(7$_{4,3}$-6$_{4,3}$) A$^-$ ($\nu_t$=1) and
HC$_3$N(37-36)($\nu_7$=1) integrated emission (blue and green
contours, respectively) and the rotational integrated emission of
CH$_3$CN(12$_3$-11$_3$) (single grey contour) toward the Orion KL
region. The white contours are from 40\% to 90\% with steps of 5\% of
the peak of the continuum emission (4.5 Jy beam$^{-1}$). The blue
contours are from 30\% to 90\% with steps of 10\% of the peak of the
line emission,  37 Jy beam$^{-1}$ km s$^{-1}$. The
integrated velocity range of the CH$_3$OH(7$_{4,3}$-6$_{4,3}$) A$^-$
($\nu_t$=2) is from $-$5 to $+$15 km s$^{-1}$. The green
contours are from 25\% to 90\% with steps of 12\% of the peak of the
line emission (170 Jy beam$^{-1}$ km s$^{-1}$).  The integrated
velocity range of HC$_3$N(37-36)($\nu_7$=1) is from $-$12 to $+$19 km
s$^{-1}$.  The single grey contour is 34\% of the peak of the line
emission (586 Jy beam$^{-1}$ km s$^{-1}$).  The integrated velocity
range of CH$_3$CN(12$_3$-11$_3$) is from $-$10 to $+$26 km
s$^{-1}$. 
Black circles and pink square as in previous Figures.
The blue and red arrows at the position of Source
{\it I} show the approximate orientation of the thermal and SiO maser
outflow emanating from this object \citep{Plambecketal2009,
Matt2010}. The white open circle marks the position of the radio
source {\it D}
\citep{Zapataetal2004a, Gomezetal2005}.}
\label{Fig6} 
\end{figure*}

\subsection{The relation between the centimeter,  
submillimeter, infrared, and molecular line emission}

An overlay of the HC$_3$N(37-36)($\nu_7$=1), the
CH$_3$OH(7$_{4,3}$-6$_{4,3}$) A$^-$ ($\nu_t$=2), and the
submillimeter
continuum emission on the 11.7 $\mu$m infrared emission from \citet{smith2005}
the Orion KL region is displayed in Figure 6.
Included in this image are the positions of the four
radio sources {\it BN}, {\it I}, {\it n}, and {\it D}
\citep{Zapataetal2004a,Gomezetal2005,rod2009}. The map shows
a lack of correspondence between  the centimeter, submillimeter,
and mid-infrared continuum sources. This poor coincidence suggests
that they might be of different nature.  We found three groups of
similar sources in the Orion KL region which we describe as follows:
\begin{itemize}

\item
The first group is formed by the compact millimeter and submillimeter
sources SMM1, SMM2 and SMM3.  These sources seem to be deeply embedded
in the Orion Molecular Cloud, with no strong mid- and near-infrared
nor centimeter emission.  They appear to be optically thick,
dusty compact objects as suggested by their large positive
spectral indices, see Table 2. 
As revealed by our observations they do not show hot core activity.

\item
A second group consists of the infrared and centimeter
sources {\it BN}, {\it I} and {\it n}. These sources have only very
faint mm or submm emission associated with them and also do not show any of the hot
molecular emission typically characterizing hot cores.

\item
A third group is formed by the extended infrared sources with
vibrationally/torsionally excited molecular emission. This
emission is related to only the
infrared objects IRS2a-d \citep[][]{dou1993,gen1998}.  This type of
source does not have any centimeter or submillimeter counterparts,
suggesting it to not be self-luminous, as first proposed
by \citet{men1995}. Maybe this thermal infrared and molecular
emission was created by shocks from the explosive disintegration
of the stellar system.
\end{itemize}

As the mid-infrared shows a good correspondence with the hot molecular
gas traced by the vibrationally/torsionally excited emission, it
would not necessarily be reprocessed emission escaping
through inhomogeneities in the dense material but could be due to
heating by shock compression that destroyed the dust grains. 

The reason that vibrationally/torsionally excited lines are observed only
towards IRS2 might be that only this spot has density or
temperature high enough to be sufficiently excited. In Figure 1 one can see that
towards the position of this emission the highly extincted and dense
Extended Ridge is located.  Molecular lines that trace colder gas, such
as NH$_3$(4,4), are not present where the mid-infrared and
vibrationally/torsionally excited molecular emission is located, see
Figure 5 of \citet{Co2006}.

We note that the Submillimeter source "Compact Ridge" (SMM1) seems to be
surrounded with mid-infrared emission oriented in approximately the direction of the
dynamical origin but without associated hot molecular gas. 

It is interesting to mention that most of the eastern OH maser spots
mapped by \citet{Co2006} show good coincidence with the
vibrationally/torsionally excited emission reported here. Moreover, part
of the OH maser emission seems to cover the submillimeter source SMM3
just as the vibrationally/torsionally excited emission
does. This further argues in favor of the infrared emission being
produced by shocks that heated the dust.

\subsection{What is Source I?}

The presence and peculiar properties of radio source {\it I} add
substantially to the complexity of the KL region's appearance. One
might think that the molecular outflow from {\it I} \citep{Plambecketal2009}
is heating the Orion KL Hot Core because it lies along the same
orientation as the vibrationally/torsionally excited lines.
However, if it were one would
expect the largest linewidths in the moment two maps (Figure 5) to be
found at the position of Source {\it I}. This is not observed.

This wide angle flow from {\it I}, whose origin is possibly an expanding,
rotating thick disk traced by SiO maser emission \citep{Matthews2010},
does not appear to be dynamically coupled to any other phenomenon in
the KL region. The observed radio emission from {\it I} is consistent with
an early type B star (B0--B1) ionizing the inner regions of the disk
\citep{Reid2007}.

As discussed by \citet{Ball05} wide angle outflows -- expanding disks
-- are a natural consequence of stellar mergers. A merger of a 20
$M_\odot$ star swallowing a 1 $M_\odot$ object will release about
$3\times10^{48}$~ergs of potential energy, more than required to power
the KL region. This line of argument could suggest source {\it I} to be the end
result of a merger. The reason for {\it BN}, {\it n}, and {\it I} all moving away from
each other after the merger event should be investigated by numerical
simulations.

It is difficult to estimate an exact dissipation timescale for the
energy released on the dynamical non-hierarchical disintegration.
However, if the central ``hole'' found on the CO(2-1) filaments
\citep{Zapataetal2009} was generated because of the molecular gas on
the outflow is cooling down, we estimated a the time of
  dissipation for the outflow on the order of five thousand
  years.

\subsection{Heating of Orion KL Hot Core by an explosive flow}

The facts that we do not find the hot molecular gas emission to be
associated with any self-luminous submillimeter, radio or infrared
source, and that this emission is restricted to the northeast edge of the
heart-shaped structure suggest that the heating source of the Orion KL
Hot Core is external.  Since HC$_3$N(37-36)($\nu_7$=1) and
CH$_3$OH(7$_{4,3}$-6$_{4,3}$) A$^-$ ($\nu_t$=1) seem to form part of a shell
around the submillimeter source SMM3, and since this shell points in
the direction of the dynamical center, the heating
source seems to be closely linked to that center.

Furthermore the unique absence of CO
filamentary flow structures or ``fingers'' from the area behind the
Orion KL Hot Core (behind relative to the outflow center) indicates that a
dense zone of the Extended Ridge might there have impeded the expansion of
such filaments.  Notably, the structures revealed by our
HC$_3$N(37-36)($\nu_7$=1) and CH$_3$OH(7$_{4,3}$-6$_{4,3}$) A$^-$
($\nu_t$=1) maps of the gas kinematics within the Hot
Core show trajectories emerging from the dynamical center.

It is important to realize that each individual CO filament obeyed a
Hubble-type velocity, {\it i.e.} its velocity increases linearly with
distance from the explosive origin \citep{Zapataetal2009}, in other
words, during the explosive event some 500 years ago material of
vastly different velocities was ejected simultaneously and has been
traveled out to different correspondingly distances. The material that
is just now arriving at the Hot Core condensation is the one that now
excites the energetic vibrations observed. We do not, therefore,
require decay times of this shock excitation to be on the order of
hundreds of years; rather, the emission we observe now is caused by
streams of matter impinging at present, not in the ``distant'' past
($\sim$ 500 years), such that relevant decay times may be very much
shorter indeed.

\citet{Kau1998} argued the Orion KL Hot Core to more likely be heated
by stars embedded within the core rather than powered from outside
because of the core's large column densities, and warm temperatures.
However, such a larger column densities and warm temperatures could be
generated by fast and energetic shocks arriving now to the core as
discussed above. It is thus not necessity of having an internal warm
stellar source to heat the Orion-KL Hot Core as proposed in
\citet{Kau1998}.

From the internally heated hot core picture one then may expected
found the hottest molecular gas associated with the central massive
protostellar source where the high temperatures and column density
reside, and can excite such molecules. This is not the case for the
Orion KL Hot Core. We found that none of the hot molecular tracers
presented here are found in association with a self-luminous
source. Moreover, the hot gas is found in an edge of the region traced
by cold and/or warm gas from other molecules (e.g. the
CH$_3$CN(12$_3$-11$_3$)), strongly suggesting external heating.  The
molecular emission with low excitation temperatures and/or critical
densities could show a very erratic behaviour for different species,
and could not pinpoint the true exciting source as already observed in
the Orion KL Hot core \citep{Wil2000,bla1996, Wri1996}, and in some
other hot cores \citep{bro2007,Moo2007}.  
  
\subsection{Consequences of KL not being a typical hot core}

In principle, there should be observable differences between the KL
region and a typical hot molecular core. For one, the chemistry in KL
should be shock driven with much higher temperatures than the
chemistry in a typical hot core. In typical cores, heating the dust
grains to just about a hundred K
suffices to evaporate complex molecules from their
grain surfaces, producing the high observed abundances of organics
which can be orders of magnitude higher than in cold molecular cloud
material \citep[see, e.g., ][]{vdTak2000, Herbst2009}.

Strong shocks that evaporate the dust grains, realising 
many molecules into the gas phase, could be better generated by low velocity 
($\sim$ 50 km $^{-1}$) C-shocks mediated grain sputtering, which seems 
to be more effective that the grain-grain collisions or even J-Shocks 
\citep{flo1995,may2000}.    

In a fast shock like the one driven by the explosive event in the KL
region, on the other hand, most molecules are probably dissociated. 
Detailed models of the ensuing chemical evolution at these much higher 
temperatures do not exist at present. In particular, it is
unknown up to what stage the evolution can proceed  on
the short time scale involved, some 500 years.

Another relevant finding is the missing class II methanol maser
in KL. Toward many hot cores associated with high mass protostars,
maser emission in the strongest class II maser line at 6.7 GHz is
observed, a.o., in Orion S \citep{Voronkov2005}. However, toward the
KL region these authors only find a feature whose line width, strength
and distribution is consistent with values found for thermal
(non-maser) emission from numerous other molecular lines. The presence
of a class II methanol maser is a sufficient, but not a necessary
condition for a source being a hot core; it is
presently unclear what percentage of hot cores feature such a maser, but probably
most do \citep{Ellingsen2006}. We find the absence of such
a maser from the KL region remarkable.

In contrast, about a dozen compact regions showing strong maser
emission in the 25 GHz maser lines, the classical class I methanol
maser transitions, are distributed all over the KL region.
Whether or not these masers have any 
relation to the explosive event (or to any other  phenomenon in the region) 
is presently completely unclear.

\section{Conclusions}

We observed and analyzed the submillimeter and millimeter
torsionally/vibrationally highly excited lines
CH$_3$OH(7$_{4,3}$-6$_{4,3}$) A$^-$ ($\nu_t$=2),
CH$_3$OH(7$_{4,3}$-6$_{4,3}$) A$^-$ ($\nu_t$=1),
HC$_3$N(37-36)($\nu_7$=1), SO$_2$(21$_{2,20}$-21$_{1,21}$)($\nu_2$=1),
as well as CH$_3$CN(12$_k$-11$_k$) with $k$=$3,6,9$ and the continuum emissions
at 870 and 1300 $\mu$m, from the Orion KL region in an attempt to
clarify the nature of the hot molecular ``core'' located in this
region. Our main findings are as follows:

\begin{itemize}

\item   The system of high velocity CO filaments
or ``fingers'' that is believed to have originated in an
explosive stellar merger event some 500 years ago is being
blocked in its southwest quadrant by the intervening Orion KL Hot Core.
The area behind the Hot Core is uniquely devoid of such filaments.
An H$_2$ image of the region
 likewise shows the northwestern fingers to be much stronger and better
 defined than southern ones;

\item  The low excitation CH$_3$CN(12$_3$-11$_3$) emission shows the typical 
heart-shaped structure mapped in many molecules; indentation of the heart-shaped
structure is centered close to the dynamical origin of the explosion;

\item The torsionally/vibrationally excited lines and CH$_3$CN(12$_9$-11$_9$), 
 all of which are supposed to trace hot and dense molecular gas, are
 located exclusively toward the northeast lobe of the heart-shaped structure, {\it
 i. e.} toward the densest and most highly obscured parts of the
 Extended Ridge.  The HC$_3$N(37-36)($\nu_7$=1) and
 CH$_3$OH(7$_{4,3}$-6$_{3,3}$) A$^-$ ($\nu_t$=1) lines appear to form
 a shell around the strongest compact submillimeter source, SMM3,
 and to point toward the dynamical origin of the CO filaments;

\item The CH$_3$CN(12$_3$-11$_3$) and HC$_3$N(37-36)($\nu_7$=1) maps of 
 the kinematics of the molecular gas within the Hot Core
 reveal filament-like structures that likewise point toward that dynamical
 origin. The peculiar turbulent velocity field of the Orion KL Hot
 Core is possibly the result of gas moving in different directions due
 to the explosive event;

\item The peaks of the emission from 
 CH$_3$CN(12$_9$-11$_9$), SO$_2$(21$_{2,20}$-21$_{1,21}$)($\nu_2$=1),
 and HC$_3$N(37-36)($\nu_7$=1) are all well concentrated in the southeast
 filamentary structure given by the moment two maps. This
suggests a close relationship between the excitation of
 these three lines and the filamentary structure;

\item Only three compact submillimeter sources 
 that are counterparts of millimeter sources already reported
 in the literature, SMM1, SMM2, and SMM3, could be found. All three
 sources show steep spectral indices suggesting optically thick dust
 emission;

\item The hottest molecular emission coincides well with a chain of
 mid-infrared (IRS2a-d) and OH maser emissions, suggesting that part
 of the infrared emission of the Orion KL region might be generated by
 strong shocks compressing the gas and evaporating the dust grains that
 realesed many molecules in the gas phase.

\item There are at least three types of objects in the Orion KL region: 
     The radio continuum sources with or without infrared emission and
     with no hot core activity ({\it BN}, {\it n}, and {\it I}) that formed a stellar
     group in the past; the submillimeter sources with no mid-infrared
     emission at all nor any hot core activity (SMM1, SMM2, and SMM3);
     and the extended mid-infrared continuum emission with or without
     associated molecular emission that probably is generated by
     strong shocks due to the explosive disintegration.

\end{itemize}

The various millimeter and submillimeter SMA observations suggest that
the Orion KL Hot Core is being heated by the explosive flow associated
with the disintegration of a massive young stellar system that is evidenced by
the three runaway objects {\it BN}, {\it n} and {\it I}. 
This hypothesis can explain most of the peculiar features observed 
in the Orion KL Hot Core such as its
strange velocity fields and peculiar morphology. However, additional 
theoretical and observational studies are required to test 
this new heating scenario. 

\begin{acknowledgements}

    We are very grateful to Nathan Smith and John Bally for having provided the 11.7
    $\mu$m infrared and the H$_2$ images. We would like also to thank to the anonymous 
    referee and Malcolm Walmsley for the detail comments and suggestions to improve this study.

\end{acknowledgements}

\bibliographystyle{aa}
\bibliography{biblio3}

\end{document}